\documentclass[letterpaper,12pt]{article}   
\usepackage{osajnl2} 
\usepackage[]{hyperref} 
\usepackage[doublespacing]{setspace}
\usepackage{graphicx}
\begin{document}

\title{Controllable ultra-broadband slow light in a warm Rubidium vapor}


\author{  Rui Zhang,$^{1,3}$ Joel A. Greenberg,$^{1,3,*}$ Martin C. Fischer,$^{2}$ and Daniel J. Gauthier$^{1}$}
\address{$^1$ Department of Physics and the Fitzpatrick Institute for Photonics, Duke University, Durham, NC 27708, USA}
\address{$^2$ Department of Chemistry and the Fitzpatrick Institute for Photonics, Duke University, Durham, NC 27708, USA}
\address{$^3$ The first two authors contributed equally to this work}
\address{$^*$ Corresponding author: jag27@phy.duke.edu}

\begin{abstract} We study ultra-broadband slow light in a warm Rubidium vapor cell.  By working between the D1 and D2 transitions, we find a several-nm window centered at 788.4 nm in which the group index is highly uniform and the absorption is small ($<$1\%).  We demonstrate that we can control the group delay by varying the temperature of the cell, and observe a tunable fractional delay of 18 for pulses as short as 250 fs (6.9 nm bandwidth) with a fractional broadening of only 0.65 and a power leakage of $55\%$.  We find that a simple theoretical model is in excellent agreement with the experimental results.  Using this model, we discuss the impact of the pulse's spectral characteristics on the distortion it incurs during propagation through the vapor.   
\end{abstract}


\maketitle 

\section{Introduction}
Controlling the propagation velocity of optical pulses in dispersive media is important both at the level of basic and applied science \cite{boydsl,boyd09}.  Fundamental studies of entanglement \cite{howell08}, information velocity \cite{stenner03}, and optical precursors \cite{jeong06} rely on the fact that one can decrease (increase) the group velocity of light $v_g=c/n_g$ in regions of normal (anomalous) dispersion to produce slow (fast) light (here $c$ is the speed of light in vacuum and $n_g$ is the group index). In recent years, a growing number of technological applications of slow-light techniques have been investigated, including all-optical buffering and routing, optical memories, and data synchronization \cite{gauthier05,gaeta06}.   While many slow light approaches make use of optical resonances  \cite{hau99, bigelow03, okawachi05} to provide large values of the group delay $t_D$ with small absorption, the spectral region over which the slow light occurs is generally narrow. Because the spectral width of the slow light region limits the minimum temporal duration $T_0$ of a pulse that can be delayed without significant distortion, it sets an upper bound on the maximum achievable data rate and fractional pulse delay ($f_D=t_D/T_0$) \cite{boyd05}.  In addition, some applications require tunable delay lines that can retard the arrival of ultrashort optical pulses ($T_0\sim$100 fs) by several hundred fs.  Thus, it is important to investigate techniques for realizing ultra-broadband slow light materials where the achievable group delay can greatly exceed the injected pulse width without causing significant pulse distortion.  

Previous methods for realizing broadband, distortion-free slow-light media include coherent control of the optical properties of a material and the design of novel materials.  As an example of an approach involving coherent control, Sharping \textit{et al.} measured $f_D=0.85$ for a 430-fs-long pulse via stimulated Raman scattering \cite{sharping05}. More recently, conversion/dispersion techniques in dispersive optical fibers have demonstrated $f_D\sim20,000$ for 2.6-ps-long pulses \cite{kurosu09} and $f_D=10$ for 370-fs-long pulses \cite{pesala09}.  Using a materials-based approach, Gan \textit{et al.} \cite{gan11} employed solid-state dispersion engineering via plasmonic structures to demonstrate fixed slow light delays over several hundred nanometers in the visible range.

In this work, we make use of the region of uniform group index between a pair of absorbing resonances to realize low-distortion, broadband slow light.  While the properties of this double resonance system have been studied previously in the context of stimulated Brillouin scattering in optical fibers  \cite{zhu06} and the hyperfine structure of atomic vapors \cite{howell06,shak08,vanner08,anderson10,camacho07}, the usable bandwidth is limited to approximately 1 GHz.  To greatly enhance the bandwidth of the slow light region, we follow the approach of Broadbent \textit{et al.} \cite{howell08} and make use of the region of normal dispersion between the fine structure absorption doublet in a warm Rubidium vapor.  Here, we focus on characterizing experimentally the wavelength and temperature dependence of $n_g$ between the D1 and D2 lines in Rubidium (comprising a splitting of 7 THz or 14 nm centered at 787 nm) as well as studying the propagation dynamics of ultrashort pulses propagating through the vapor.  Using a simple, first-principles model, we obtain excellent agreement between our experimental results and theoretical predictions. We observe a maximum value of $n_g$ equal to 1.03, which is uniform (less than 50\% variation) over a $\Delta\nu=3$ THz ($\Delta\lambda=6$ nm) bandwidth with a nearly exponential sensitivity to temperature.  Also, we measure the distortion of pulses transmitted through the vapor, and show that pulses as short as 250 fs can undergo fractional delays of almost 20 while suffering a fractional broadening $f_B$ of less than 0.65. 

The paper is organized as follows.  The next section outlines the theoretical model we use to describe the transmission of light through the vapor cell.   Section \ref{sec:exp} describes the experimental setups we use to characterize the slow light medium, Sec. \ref{sec:disc} compares the experimental and theoretical results, and Sec. \ref{sec:conc} concludes the paper and gives possible future applications of our results.

\section{Theory}
\label{sec:theory}

Atomic vapors are a well understood material whose optical properties can be precisely predicted and controlled. In this work, the optical frequency is so far detuned (by $>3$ THz) from either the D1 or D2 resonances of Rubidium that we can treat each atom as consisting of multiple two-level systems (\textit{i.e.}, ignore coherences between the levels considered).  Furthermore, because of this large detuning, we can ignore Doppler broadening and approximate the Voigt profile of the resonance (obtained via convolution of the Gaussian profile due to Doppler broadening and the Lorentzian profile due to collisional and natural broadening) as a Lorentzian \cite{camacho07}.   Thus, we can describe the atomic susceptibility $\chi_{full}(\omega)$ experienced by a weak optical field as it propagates through a vapor cell containing Rubidium vapor as a sum of Lorentzian functions weighted appropriately by the various transition strengths (see  Eq. (8) in Ref. \cite{howell08}). Because we work so far from the atomic resonances, the details of the underlying hyperfine structure are not important.  This allows us to reduce the full model of the atomic susceptibility to an effective double-Lorentzian susceptibility for the D1 and D2 lines given as
\begin{equation}
\label{eq:redchi}
\chi_{eff}(\omega)=-N(T)\left( \frac{s_1}{\omega-\omega_1+i\gamma_1}+\frac{s_2}{\omega-\omega_2+i\gamma_2} \right),
\end{equation}
where $s_1=2.25\times10^{-13}$ and $s_2=4.58\times10^{-13}$  m$^3$rad/s are effective transition strengths, $\omega_1$ and $\omega_2$ are the effective resonance frequencies, and $\gamma_1$ and $\gamma_2$ are the effective linewidths (including both natural and collisional broadening) for the D1 and D2 transitions, respectively \cite{steck}.   We model the temperature-dependent atomic density $N(T)$ via the vapor pressure relations given in Ref. \cite{nesmeyanov63} by treating the Rubidium vapor as an ideal gas. For the range of parameters considered in the remainder of the paper, the normalized deviation  between $\chi_{full}$ and $\chi_{eff}$ (defined as $|(\chi_{full}-\chi_{eff})/(\chi_{full}+\chi_{eff})|$) is less than $5\times10^{-3}$, thus making their predictions nearly indistinguishable.

	The complex index of refraction is related to the susceptibility as $n=n'+in''\approx1+\chi_{eff}/2$ (since $\chi_{eff}\ll1$ for the regime studied here).  This allows us to compute the absorptive optical depth as $\alpha L=2 \omega L n''/c$, where $L$ is the interaction length and $\alpha$ is the intensity absorption coefficient.  Because we assume only weak optical fields propagating through the vapor, we describe the fractional  absorption of light using Beer's Law as $1-I_{out}/I_{in}=1-exp(-\alpha L)$.  The group index is found through the relation 
$n_g(\omega')=n'+\omega  dn'/d\omega|_{\omega=\omega'}$, which allows us to calculate the group delay $t_D=L(n_g-1)/c$.  We note that the group velocity dispersion, given as GVD$=d(1/v_g)/d\omega$, is zero at $\lambda_c=788.4$ nm (at the center of gravity between the D1 and D2 lines).
	
To study quantitatively pulse propagation in our system, we  simulate numerically the propagation of an incident pulse with a field envelope $E(0,t)$ and intensity full width at half maximum (FWHM)) $T_0$.   After traveling a distance $z$ in the vapor, the field envelope of the pulse is 
\begin{equation}
E(z,t)= \mathcal{F}^{-1} \left(\mathcal{F}(E(0,t)) H(\omega)\right)
\end{equation}
where $\mathcal{F}$ represents the Fourier transform and $H(\omega)=exp(i n(\omega) \omega/c)$ is the transfer function of the vapor.  The fractional delay and broadening are calculated, respectively, as $f_D=t_D/T_0$ and $f_B=(T-T_0)/T_0$, where $t_D$ is the time delay of the pulse peak and $T$ is the FWHM of the pulse when it exits the vapor.

\section{Experimental Details}
\label{sec:exp}

\begin{figure}
 
  \begin{center}
  \includegraphics[width=6. in]{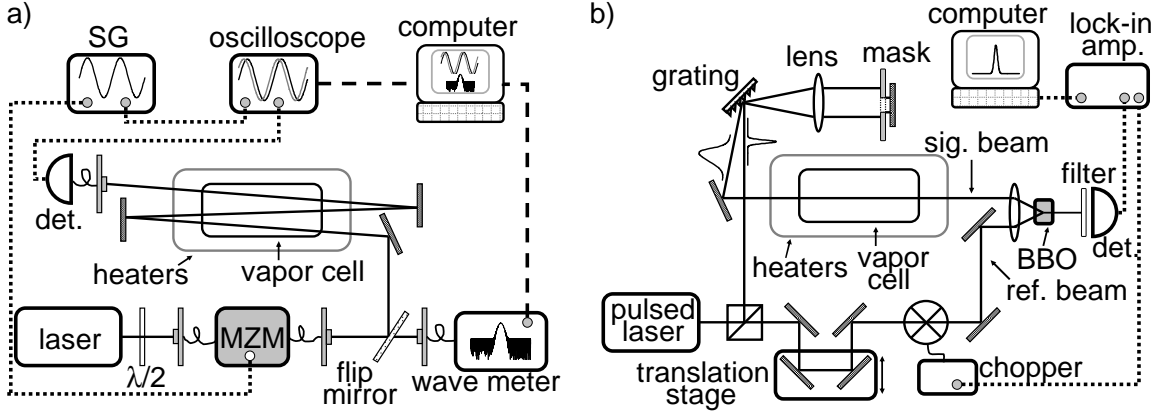}
  \end{center}
  \caption{Schematics of the experimental setups.  a) A weak beam is modulated by a MZM, passes three times through a heated Rubidium vapor cell (Triad Technologies, TT-RB-75-V-Q-CW), and is detected by a fast photoreceiver (New Focus 3051). b) An ultrashort pulse passes once through the vapor cell, overlaps with a tunable reference pulse in a BBO crystal, and the resulting sum-frequency generated light is recorded.}
  \label{fig:setup}
\end{figure}

In this paper, we carry out two distinct but related experiments in order to fully characterize the optical response of the Rubidium vapor.  In the first experiment, we use a monochromatic source to measure directly the frequency-dependent susceptibility of the vapor (see Fig. \ref{fig:setup} a)).  We use an external cavity diode laser (Toptica, DL Pro) to generate continuous wave light with a narrow linewidth ($<1$ MHz) that we  tune across the D1 and D2 resonances (from 772-796 nm).  We monitor the frequency of the light via a wavemeter with 2 MHz resolution (WS Ultimate Precision, High Finesse).  In order to modulate the amplitude of the optical signal, we couple the light into a Mach-Zehnder modulator (MZM) via a free-space-to-fiber port, and operate the MZM in the linear regime. We note that the MZM introduces dispersion, but we explicitly account for this in our analysis.  An RF signal generator (SG, HP 70340A) drives the MZM at frequencies of up to 7 GHz (limited by the SG).  We then couple the modulated optical signal back to free space where it interacts with a heated 7 cm vapor cell containing natural Rb in a multi-pass geometry.  After three passes through the cell ($L=21$ cm), we couple the light  into an optical fiber, detect the signal, and record the resulting waveforms with a high-speed oscilloscope (Tektronix 11801B).  

In the second experiment, we inject ultrashort laser pulses into the vapor and measure the shape and absolute delay of the outgoing pulses (see Fig. \ref{fig:setup} b)).  We use a mode-locked Ti:Sapphire laser (Spectra-Physics, Tsunami) to generate ultra-short pulses ($\sim$ 55 fs) centered between the D1 and D2 resonances, then split the laser output into a reference and signal beam.  The reference beam passes through a tunable optical delay line (used to scan the temporal delay between the signal and reference beam) and chopper, and is then focused into a beta barium borate (BBO) crystal.  To tailor the frequency spectrum of the signal beam, we send it to a grating, use a rectangular mask to remove unwanted frequency components, and then re-compress the signal beam by sending it back to the grating.  In this way, we  produce nearly Fourier-limited, sinc-shaped pulses with a width between $\sim$100 fs and 2 ps by adjusting the spatial width of the mask.  We then pass the signal beam through the vapor cell and focus the transmitted beam down onto the BBO crystal.  When the signal and reference beams are spatially and temporally overlapped in the BBO crystal, sum-frequency generated light is produced.  By varying the relative delay between the signal and reference beam, we measure the cross-correlation between the two beams via a photodiode and lock-in amplifier.

To control the temperature of the Rubidium vapor, we place the vapor cell in a cylindrical enclosure with anti-reflection coated windows.  We heat the cell using ceramic heaters and insulate using fiberglass around the body of the cell; we do not directly heat the vapor cell cold finger.  We independently monitor the temperature at the cold finger and the front and back faces via K-type thermocouples, and ensure that the cold finger is always 20 $^\circ$C colder than the faces of the cell so that Rubidium does not condense on the cell walls.  This setup allows us to vary the cold finger temperature between $18$ and 350 $^\circ$C.  Finally, to account for any temperature-dependent changes in the index of refraction of the cell or air in the heated enclosure, we repeat the experiments for an empty vapor cell.  Thus, our measurements depend only on the optical properties of the Rubudium vapor.

\section{Results and discussion}
\label{sec:disc}

\begin{figure}[h]
  
  \begin{center}
  \includegraphics[width=6 in]{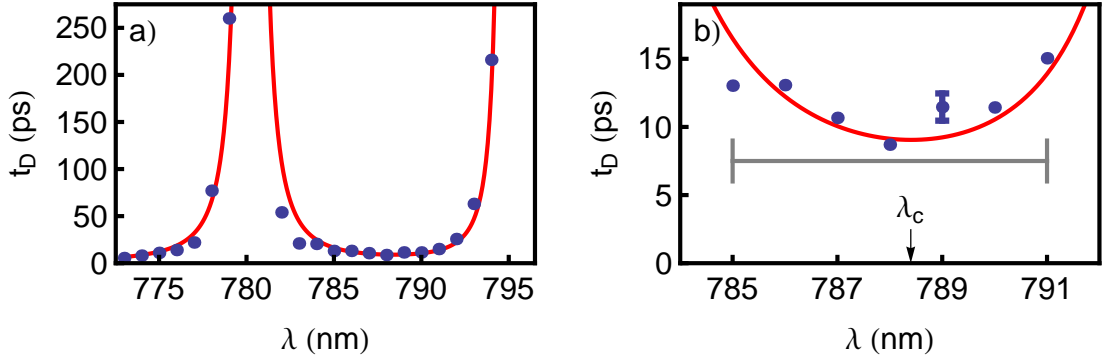}
  \end{center}
  \caption{(Color online) Experimental (points) and theoretical (solid lines) values of the wavelength-dependent slow light delay measured  a) across the entire region around the D1 and D2 resonances and b) in the region of uniform group index between the resonances.  The grey horizontal bar indicates the 6 nm region of minimal group index variation centered on $\lambda_c$.  The data correspond to a probe beam passing three times through a 7 cm vapor cell at a temperature of 280 $^\circ$C. The error is the same for all points and is due to uncertainties in the vapor temperature measurement. }
  \label{fig:delay}
\end{figure}

To validate our model, we begin by comparing the theoretically-predicted and experimentally-measured group delays.   Figure \ref{fig:delay} a) shows the wavelength dependence of the group delay across the D1 and D2 transitions (measured via a monochromatic source) for a cell temperature of 280 $^\circ$C.  In agreement with the model, we measure large ($\sim$250 ps) and rapidly-varying delays near the resonances, and smaller ($\sim$10 ps) but uniform delays between the resonances.  To highlight the uniform broadband slow light region, Fig. \ref{fig:delay} b) shows an expanded view of the group delay between 784 and 792 nm. The group index varies by less than $50\%$ over 6 nm and has a minimum at $\lambda_c$ (\textit{i.e.}, the point where GVD$=0$).

\begin{figure}[h]
  
  \begin{center}
  \includegraphics[width=6 in]{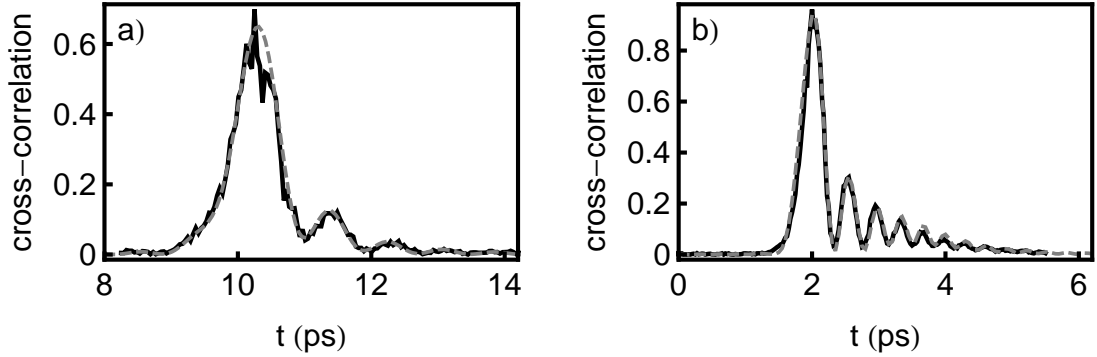}
  \end{center}
  \caption{(Color online) Comparison of experimental (solid line) and theoretically predicted (dashed line) results for a a) 650-fs-long pulse centered at $\lambda_c$ and a vapor temperature of 326 $^{\circ}$C and b) 250-fs-long pulse centered at 787.5 nm and a vapor temperature of  263 $^\circ$C.   }
  \label{fig:pulse}
\end{figure}

Next, we compare the experimental results for pulse propagation through the vapor with the model predictions.  Figure \ref{fig:pulse} shows the shape of pulses after they have passed through a 7 cm length of the atomic vapor.  The amplitude is obtained, as described in Sec. \ref{sec:exp}, via a cross-correlation measurement with a temporally short reference beam, and we scale the results such that a pulse propagating through vacuum has a peak amplitude of 1 and an arrival of its peak at a time $t=0$.  We find excellent agreement between the model and experimental data for 650 and 250-fs-long pulses centered at 787.5 and 788.4 nm after they have traveled through the cell heated to 326 and 263 $^\circ$C (see Fig. \ref{fig:pulse} a) and b), respectively).  After propagation through the vapor, the pulses become broader and develop trailing intensity oscillations (we quantify this distortion below).  Because we are working so far from the atomic resonances, we find that the main contribution to pulse distortion stems from dispersive effects (\textit{i.e.}, absorption-induced distortion is negligible), as described in Appendix. \ref{sec:appa}.  We also note that our system is nearly lossless; the decrease in pulse amplitude is a direct result of power redistribution rather than loss due to absorption (we calculate the absorption to be $<1\%$).  

\begin{figure}[h]
 
  \begin{center}
  \includegraphics[width=6 in]{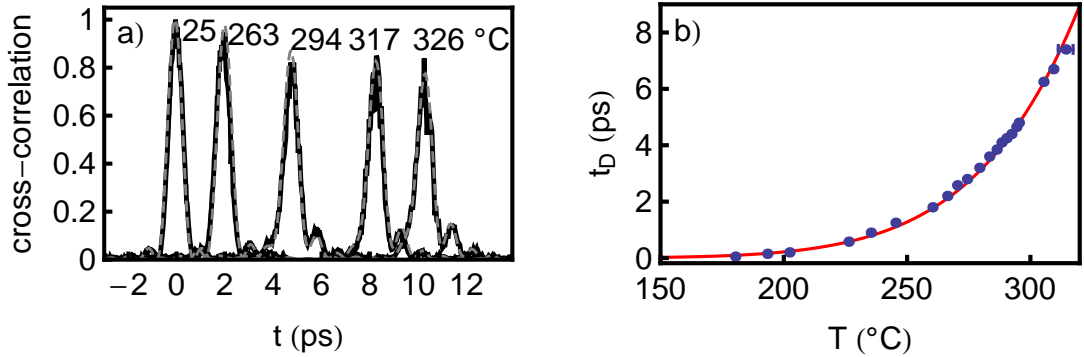}
  \end{center}
  \caption{(Color online) a) Comparison of experimental (solid line) and theoretically predicted (dashed line) results for multiple 650-fs-long pulses at 25, 263, 294, 317, 326  $^{\circ}$C (from left to right, respectively) centered at $\lambda_c$.  b) Dependence of the experimental (points) and theoretical (solid line) values of the group delay on cell temperature.  The error is the same for all points and is due to uncertainties in the measurement of the vapor temperature. }
  \label{fig:temp}
\end{figure}

We find that our model also  describes accurately pulse propagation for a broad range of temperatures.  Figure \ref{fig:temp} a) shows the output pulse shapes for identical incident 650-fs-long  pulses traveling through the vapor at various temperatures.  The group delay depends  strongly (nearly exponentially) on the temperature, as shown in Figure \ref{fig:temp} b), which allows us to easily tune the group delay over almost 10 ps in a 7 cm cell.  

\begin{figure}[h]
  
  \begin{center}
  \includegraphics[width=3 in]{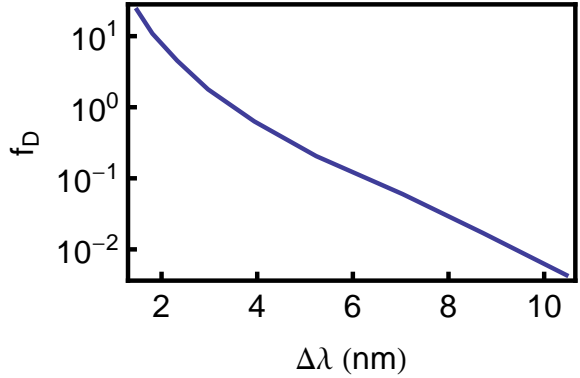}
  \end{center}
  \caption{(Color online) Theoretically-predicted best achievable fractional delay vs. input pulse bandwidth (centered at $\lambda_c$) subject to the constraint that $D_a,D_p<$0.05, which corresponds to negligible distortion (\textit{i.e.}, just noticeable upon visual inspection). }
  \label{fig:bandwidth}
\end{figure}

Knowing the full susceptibility of the vapor gives us complete knowledge of the vapor's effective transfer function $H(\omega)$ and allows us to quantify the pulse distortion.  While the distortion tolerances for a system can be highly specific to the intended application, we take two different approaches to place our work in context with other studies.  First, we calculate the deviation of the Rubidium vapor's transfer function from an ideal one for a given bandwidth of the incident pulse.  The ideal transfer function, which induces no distortions, is defined as  $H_{ideal}(\omega)=H_0 exp(i \omega t_p)$, where $H_0$ is a constant amplitude and $t_p$ is the total propagation time of the pulse. Using the concept of amplitude and phase distortion ($D_a$ and $D_p$, respectively) defined in Ref. \cite{stenner05}, we find that phase distortion is the limiting quantity for realizing low-distortion delays over large bandwidths ($D_a<10^{-3}$ for the regime studied in this paper).  We note that this measure of distortion is particularly relevant to our system because we use pulses with spectra that are well-approximated by a rectangular function.  By defining a maximum-allowed distortion $D^{max}$, we  calculate the largest achievable fractional delay for a given spectral bandwidth subject to the constraint that $D_a,D_p<D^{max}$ (see Fig. \ref{fig:bandwidth}).  Thus, we find that one can realize $f_D>1$ over a bandwidth of several nanometers while incurring negligible distortion (\textit{i.e.}, $D^{max}<0.05$).

\begin{figure}[h]
  
  \begin{center}
  \includegraphics[width=6 in]{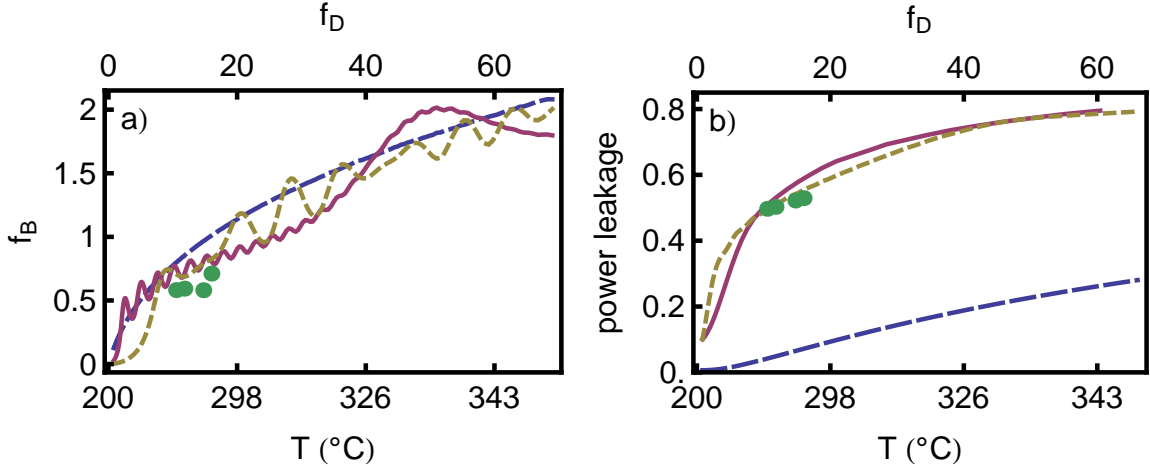}
  \end{center}
  \caption{(Color online) a) Fractional broadening and b) power leakage versus vapor temperature (as well as the corresponding fractional delay) for a 250-fs-long pulse propagating through a 7 cm cell.  The long dashed line, short dashed line, and solid line correspond to theoretical results for a pulse with a Gaussian spectrum centered at $\lambda_c$=788.4 nm, a rectangular spectrum centered at $\lambda_c$, and a rectangular spectrum centered at 787 nm, respectively.  The points are experimental data for a pulse with a rectangular spectrum centered at 787 nm.}
  \label{fig:fbpow}
\end{figure}

This metric does an excellent job of quantifying the pulse distortion, but it is extremely conservative in that the entire pulse shape is considered.  While this may be appropriate for applications where one needs to maintain the details of the pulse's temporal profile (such as for biomedical imaging using pulse-shaping techniques), many applications are less sensitive to the exact pulse shape.  To quantify pulse distortion in an alternative manner, we adopt two additional distortion metrics: the fractional broadening of the pulse and the power leakage (defined below).   Figure \ref{fig:fbpow} a) shows the fractional broadening as a function of the vapor temperature for a 250-fs-long pulse. We observe $f_D=18$ for $f_B=0.65$ (corresponding to a vapor temperature of 294  $^{\circ}$C), in good agreement with our theoretical predictions. To demonstrate the impact of the pulse shape and central wavelength, we show theoretical results for the case where the incident pulse has a rectangular spectrum centered at $\lambda_c$ and 787 nm, as well as a Gaussian spectrum centered at $\lambda_c$. While the fractional broadening of the sinc pulses generally follows the trend set by the Gaussian pulse, it is clear that careful selection of pulse shape and central wavelength are crucial for minimizing broadening. 

To go beyond pulse broadening and quantify the potential impact of the propagation-induced intensity oscillations, we compute the power leakage, which is defined as the fraction of total pulse power extending beyond a temporal window encompassing the main part of the pulse.  To calculate the power leakage, we first establish a temporal window to define the main pulse (this can be thought of as a channel in the context of information theoretic applications \cite{neifeld08}).  For our sinc pulses, we choose the window to be the time between the first minima surrounding the main pulse peak (before it passes through the vapor).  We then adjust the center of this window until it contains the largest fraction of transmitted pulse power;  the fraction of total pulse power outside this window is the leaked power. By using this metric, we gain insight into how higher order distortion contributes to spreading of the pulse beyond its initial channel. 

 Figure \ref{fig:fbpow} b) shows the power leakage for a 250-fs-long pulse with several different spectral profiles.   We experimentally observe a power leakage of $65\%$ for a fractional delay of 18 (corresponding to a vapor temperature of 294 $^{\circ}$C).   We note that, for an undistorted sinc pulse, approximately $10\%$ of the pulse power extends beyond the main pulse. Thus, although the broadening of the Gaussian and sinc pulses are comparable, the power leakage for the Gaussian pulse is almost $45\%$ less than that of the sinc pulse due to the fact that the spectral bandwidth of the Gaussian pulse is less than that of the sinc-shaped pulse for identical FWHM temporal pulse widths.  Gaussian pulses therefore appear to be advantageous for limiting cross-talk between closely-spaced pulses. Furthermore, the effects of propagation on Gaussian pulses can be determined analytically, which allows for additional insight into the mechanisms responsible for distortion.

\section{Conclusions}
\label{sec:conc}
We measure the group index spectrum of a warm Rubidium vapor in a multipass geometry using a monochromatic laser source in the vicinity of the D1 and D2 lines as well as the medium's pulse response using ultrashort pulses on the order of several hundred femtoseconds.  We demonstrate that calculations involving a simple, Lorentzian frequency profile for each of the involved fine states describe well the experimental results, which allows us to use the model to make further predictions about the behavior of the system.  Also, we show that we can tune the group delay over a wide range of values by simply varying the temperature of the vapor cell.  While this method is very simple, it is rather slow.  As an alternative approach, one can apply a separate, strong saturating pulse near one of the resonance lines to optically pump atoms out of the ground state and thereby rapidly ($\sim$100's of ns) vary the group index \cite{camacho07}.  

Furthermore, this atomic-vapor-based scheme can be readily converted to a chip-based geometry \cite{yang07}.  By simply increasing the number of passes that the laser beam makes through the cell, one can either shrink the size of the required vapor cell to obtain a fixed group delay, or increase the total delay for a fixed cell length.  One can also use different atomic species to obtain slow light centered on different wavelengths.  Such a tunable, broadband slow light medium could be useful for developing more sensitive interferometers \cite{shi08}, variable-depth OCT systems with no moving parts \cite{rui10}, optical precursor experiments, or for quantum light-matter interfaces \cite{akopian11}.

 We gratefully acknowledge the financial support of the DSO Slow Light program and the Air Force Research Laboratory under contract FA8650-09-C-7932.  Also, the authors would like to thank Warren Warren for the use of his lab in carrying out the ultrashort pulse experiments.

\appendix
\section{Appendix A: Origin of pulse broadening}
\label{sec:appa}

In general, a pulse with a spectrum located between a pair of resonances will experience broadening due to the combined effects of dispersion and absorption as it propagates.  Depending on the parameters involved, limiting cases exist in which this broadening is almost completely determined by either frequency-dependent absorptive or dispersive effects.  For a pair of Lorentzian absorbing lines, as described in Eq. \ref{eq:redchi}, and a Gaussian pulse envelope, we  analytically evaluate the relative importance of these two effects.  We can calculate separately the broadening due to absorption and dispersion if we assume that the pulse is centered at the center of gravity of the transitions,  $s_1=s_2$, and $\gamma_1=\gamma_2=\gamma$  \cite{boyd05,camacho07}.  Following the approach taken by Boyd \textit{et al.} in Ref. \cite{boyd05}, we set the pulse broadening caused by each mechanism equal to a fixed value (here we choose $f_B=1$) and calculate the length of the slow light medium required to produce that amount of broadening.  We find that 
\begin{equation}
\label{eq:lad}
\frac{L_A}{L_D}=\frac{| -6+36( \omega_{21}/\gamma )^2-6 ( \omega_{21}/\gamma )^4 |}{2 \gamma T_0 
	[( \omega_{21}/\gamma )^2+1 ] [ 6 ( \omega_{21}/\gamma )^2-2 ] },
\end{equation}
where $L_{A,D}$ are the required path lengths for the absorptive and dispersion contributions and $\omega_{21}=(\omega_2-\omega_1)/2$ is half of the separation between the centers of the resonances.  Absorptive broadening is larger than (smaller than) dispersive broadening for $L_A/L_D>1$ ($L_A/L_D<1$). In general, the relative importance of absorption and dispersion depends both on the pulse width and resonance line separation normalized by the resonance linewidth.  For the case where $\omega_{21}/\gamma>>1$ (which is the case for the atomic resonances studied in this paper), Eq. \ref{eq:lad} reduces to $L_A/L_D\sim1/2 \gamma T_0$ and depends only on the pulse width relative to the atomic linewidth.  Thus, as $T_0>>\gamma$ for all of the experiments reported in this paper, dispersion is the dominant broadening mechanism.

\end{document}